\documentclass{aa}
\usepackage{graphicx}
\usepackage[varg]{txfonts}
\usepackage{hyperref}
\hypersetup{colorlinks=true, linkcolor=blue, urlcolor=blue, citecolor=blue}

\begin{document}

\title{Top-down formation of ethylene from fragmentation of superhydrogenated polycyclic aromatic hydrocarbons
\thanks{The data set associated with this work can be found under \href{https://gitlab.au.dk/tang/data-C2H4-from-fragmentation-of-HPAHs}{https://gitlab.au.dk/tang/data-C2H4-from-fragmentation-of-HPAHs}.}}

\author{Zeyuan Tang \inst{1}
\and Frederik Doktor S. Simonsen \inst{1}
\and Rijutha Jaganathan \inst{1}
\and Julianna Palotás \inst{2}
\and Jos Oomens \inst{2}
\and Liv Hornekær \inst{1}
\and Bjørk Hammer \inst{1} 
}

\institute{Center for Interstellar Catalysis, Department of Physics and Astronomy, Aarhus University, Ny Munkegade 120, Aarhus C 8000, Denmark
\\ \email{hammer@phys.au.dk}
\and Radboud University, Institute for Molecules and Materials, FELIX Laboratory, Toernooiveld 7, 6525ED Nijmegen, The Netherlands
}

\date{\today}

\abstract
{
Fragmentation is an important decay mechanism for polycyclic aromatic hydrocarbons (PAHs) under harsh interstellar conditions
and represents a possible formation pathway for small molecules such as H$_2$, C$_2$H$_2$, and C$_2$H$_4$.
}
{
Our aim is to investigate the dissociation mechanism of superhydrogenated PAHs that undergo energetic processing
and the formation pathway of small hydrocarbons.
}
{
We obtain, experimentally, the mass distribution of protonated tetrahydropyrene (C$_{16}$H$_{15}^+$, $[py+5H]^+$) and protonated hexahydropyrene (C$_{16}$H$_{17}^+$, $[py+7H]^+$)
upon collision-induced dissociation (CID).
The infrared (IR) spectra of their main fragments are recorded by infrared multiple-photon dissociation (IRMPD).
Extended tight-binding (GFN2-xTB) based molecular dynamics (MD) simulations were performed in order to provide the missing structure information for this experiment
and to identify fragmentation pathways.
The pathways for fragmentation were further investigated at a hybrid density functional theory (DFT) and dispersion-corrected level.
}
{
A strong signal for loss of 28 mass units of $[py+7H]^+$ is observed both in the CID experiment and the MD simulation,
while $[py+5H]^+$ shows a negligible signal for the product corresponding to a mass loss of 28.
The 28 mass loss from $[py+7H]^+$ is assigned to the loss of ethylene (C$_2$H$_4$) and
a good fit between the calculated and experimental IR spectrum of the resulting fragment species is obtained.
Further DFT calculations show favorable kinetic pathways for loss of C$_2$H$_4$ from hydrogenated PAH configurations involving three consecutive CH$_2$ molecular entities.
}
{
This joint experimental and theoretical investigation proposes a chemical pathway of ethylene formation from fragmentation of superhydrogenated PAHs.
This pathway is sensitive to hydrogenated edges (e.g., the degree of hydrogenation and the hydrogenated positions).
The inclusion of this pathway in astrochemical models may improve the estimated abundance of ethylene.
}

\titlerunning{C$_2$H$_4$ formation from fragmentation of HPAHs}
\authorrunning{Zeyuan Tang et al.}
\maketitle 

\section{Introduction}
Polycyclic aromatic hydrocarbons (PAHs) are abundant in space, as
their vibrational features are consistent with aromatic infrared bands
(AIBs) at 3.3, 6.7, 7.7, 8.6, and 11.3 $\mu$m which are among the
strongest emission features observed in the interstellar medium (ISM)
\citep{tielens2008}.  In the ISM, especially in the photodissociation
regions where high ultraviolet (UV) fields exist, PAHs may get
electronically excited by absorbing UV and visible photons.
The excited PAHs subsequently dissipate their excess energy via
several competing channels: ionization, infrared (IR) emission,
isomerization, and fragmentation \citep{tielens2013}.  
 PAH fragmentation is believed to play an important role in the
catalytic formation of H$_2$ \citep{rauls2008, thrower2012}, as a source
of small hydrocarbons \citep{chacko2020}, and as a part of more complex
top-down interstellar chemistry \citep{zhen2014}.  It is also an
indicator of whether or not PAHs can survive in harsh interstellar
environments \citep{reitsma2014,gatchell2015}.  Experimental techniques,
such as imaging PhotoElectron Photo-Ion
COincidence (iPEPICO) \citep{west2014,candian2018a}, infrared multiple-photon
dissociation (IRMPD) \citep{IRMPD_review,bouwman2016}, and temperature-programed desorption (TPD) \citep{thrower2012}, have been used to
investigate the fragmentation of PAHs in neutral \citep{thrower2012},
cationic \citep{castellanos2018}, and hydrogenated
form \citep{campisi2020}, providing mass-to-charge ratios of fragments.
The resulting mass spectra are usually dominated by H, 2H/H$_2$, and
C$_{2n}$H$_x$ loss \citep{jochims1994, zhen2015, rapacioli2018,
  castellanos2018, west2019, wiersma2020}.  The lack of structural
information in mass spectra of intermediate and product fragments,
however, strongly limits our understanding of the detailed processes
and pathways in PAH fragmentation.

As a way to find the missing information for experiments, static
quantum chemical calculations, such as density functional theory (DFT),
have been used to aid the interpretation of mass spectra by revealing
possible structures and mapping out the reaction network during
fragmentation.  Such a method has been successfully applied in
understanding the reaction pathways for H and H$_2$ losses during PAH
fragmentation \citep{rauls2008,thrower2012,castellanos2018}.  An evident
weakness of this static approach is the lack of information about the
mass distribution of the investigated fragments.  This can be overcome
within the Rice-Ramsperger-Kassel-Marcus theory \citep{solano2015,
  west2018}, using DFT-calculated transition state energies.  However, prior knowledge about
fragmentation channels is needed (which is often unknown) and the
dynamical process during fragmentation cannot be probed.  Another
challenge is to interpret the pathways leading to larger mass losses
than H$_2$ due to the more complex rearrangements of chemical bonds.
For H$_2$ loss, it is feasible to enumerate all possible routes of
H$_2$ formation and to calculate reaction barriers for each pathway using
DFT.  In principle, this can also be done for C$_{2n}$H$_x$
loss \citep{west2019}, but it requires so many trials and the use of chemical intuition
which might not cover all possible fragmentation pathways.
Experimental evidence for the molecular structures that formed upon fragmentation
is therefore urgently needed.

Another popular method to study fragmentation is Born-Oppenheimer
molecular dynamics (MD), which is capable of directly probing all the
relevant dynamical processes \citep{simon2017,chen2019,chen2020} and providing the mass distribution
of the fragments \citep{simon2018,rapacioli2018}.  MD covers
all possible fragmentation pathways in an unbiased way as long as
a sufficient number of runs is conducted.  Generating one theoretical mass
spectrum, however, requires enormous amounts of single point evaluations during the
MD simulation.
It is often only affordable with a semi-empirical potential \citep{bauer2016}
and for trajectories that are much shorter than the experimental time scales
established for intra-molecular vibrational redistribution (IVR) mediated fragmentation.  
This dynamical-based method is also
limited by the poor description of the chemical kinetics behind the
fragmentation, that is transition state structures and energy barriers,
while DFT or other higher level ab initio methods are better at this.

In the present work, we present a combined experimental and
theoretical approach in the study of the fragmentation of two
superhydrogenated pyrenes. The two molecules studied are
both singly charged cationic pyrene with either five additional
hydrogen atoms, $[py+5H]^+$ (4, 5, 9, 10-tetrahydropyrene) or seven
additional hydrogen atoms, $[py+7H]^+$ (1, 2, 3, 6, 7, 8-hexahydropyrene).
The experimental work was conducted as collision-induced dissociation (CID) and IRMPD studies that provide mass distributions and IR vibrational
fingerprints of the fragmentation products. The computational work is
comprised of MD simulations performed with an extended
tight-binding (GFN2-xTB) energy expression. For the fragments
identified by the MD simulations, more elaborate hybrid-DFT calculations were performed to provide reliable IR vibrational frequencies at the
harmonic approximation level, and to facilitate the discussion of
reaction energies and reaction barriers.

The paper is organized as follows.  Sect. \ref{s-exp}
describes how to obtain experimental mass spectra of $[py+5H]^+$ and
$[py+7H]^+$ and IR spectra of their fragments.  In Sect. \ref{s-Methods}
theoretical methods used in this work are briefly described, including
how to get theoretical mass spectra from MD, as well as how to
calculate energy barriers and IR spectra.  Sect. \ref{s-mass}
presents our results for yields of 28 (in atomic mass unit, amu/charge) mass loss in the
theoretical and experimental mass spectra for the two protonated pyrenes.
The confirmation of ethylene formation being assigned to 28 mass loss
is achieved in Sect. \ref{s-C2H4} by identifying the fragment structure
through IR calculations.  This is followed by Sect. \ref{s-edge} which
reveals the role of edge structures in ethylene formation.  Then
Sect. \ref{s-PAHs} presents ethylene formation on other PAHs with specially
designed edge structures.  Sect. \ref{s-astro} discusses the astronomical
implication of this work.  Finally, the paper concludes in
Sect. \ref{s-conclusion}.

\section{Methodology}
\subsection{Experimental methods} \label{s-exp}
All experiments were carried out at user station 9 of the Free Electron Lasers for
Infrared eXperiments (FELIX) facility, Radboud University, Nijmegen,
the Netherlands.
Experiments were carried out in a modified 3D quadrupole ion trap (Bruker Amazon Speed ETD).
User station 9 employs a tandem mass spectrometry
(MS/MS) setup for action gas phase spectroscopy which has been
modified to allow for direct interaction between trapped ions and
radiation from FELIX; for a detailed description, readers can refer to \citet{martens2016}.
IRMPD was
performed to measure the IR spectra of molecular ions.  The
IRMPD process happens when complex molecules are heated by
the absorption of photons whose energy becomes distributed among the
vibrational degrees of freedom of the molecules via intra-molecular
vibrational redistribution (IVR)
\citep{IRMPD_review,IRMPD_tutorial_review}.  The superhydrogenated
pyrenes (4, 5, 9, 10 - tetrahydropyrene (C$_{16}$H$_{14}$, $[py+4H]$)
($>$98\%; TCI) and 1, 2, 3, 6, 7, 8 - hexahydropyrene (C$_{16}$H$_{16}$,
$[py+6H]$) (97\%; Sigma-Aldrich)) are introduced to the systems
dissolved in toluene creating a 1 mmol
stock solution, and then diluted further in a 50\% / 50\% methanol-toluene mixture.  The ions are formed by spraying the solution into an
atmospheric pressure chemical ionization (APCI) source. The APCI source may
either form cations by removing electrons or by protonation, the
latter of which is leading to the examined HPAHs $[py+5H]^+$ and
$[py+7H]^+$.  The MS/MS setup can also produce fragments via CID in a
MS/MS step prior to exposure to FELIX, hence providing the option to
examine the mass and IR spectra of both parent and fragmented ions.
FELIX produces intense short-pulsed IR radiation in discrete steps
across the preset wavelength range of 600 - 1800 cm$^{-1}$. The
radiation is delivered in 5 - 10 $\mu$s macropulses at 10 Hz with an
approximate energy of 40 - 100 mJ/pulse and a bandwidth of 0.4 \% of
the set wavelength.  For these experiments, the IRMPD settings were
optimized yielding 2 macropulses with 3 dB attenuation.  Each macropulse
consists of a train of 6 ps long micropulses spaced by 1 ns which
drive the IRMPD.  The presented IR spectra were produced by taking the
fragment to the total ion ratio:
$\text{yield} = \sum I_{\text{fragments}}/\sum I_{\text{all ions}}$.
The wavelength was calibrated using a grating spectrometer and believed to be accurate to within $\pm$ 0.02 $\mu$m.

\subsection{Computational methods} \label{s-Methods}
The fragments of the investigated ions in this work were
experimentally generated via CID.  We assume the increased internal
energy of ions due to collisions is randomly redistributed into the
bath of vibrational degrees of freedom,  so that subsequent fragmentation
is statistical \citep{chen2014a}.
This enables the modeling of the collision-induced fragmentation merely through MD simulations at elevated temperatures
which has also been  done by \citet{koopman2021}.
The key element of MD
simulations to successfully describe statistical fragmentation is the choice of a
proper inter-atomic potential for determining energies and forces of a
given system.  The semi-empirical method PM3 has been utilized to
study the fragmentation of pristine \citep{chen2019} and functionalized 
\citep{chen2020} PAHs, while the density functional-based tight binding
(DFTB) has been used to study the competition between hydrogenation and
fragmentation by comparing experimental and theoretical mass
spectra \citep{rapacioli2018}.  It is important to consider both
accuracy and computational costs when choosing the potential for
running MD simulations.  The recently proposed extended tight-binding
method, GFN2-xTB \citep{GFN2-xTB}, appears to have an acceptable
accuracy-to-cost ratio.  The combination of MD and GFN2-xTB has
successfully reproduced the experimental mass distribution of a wide
range of organic, inorganic, and organometallic systems, while detailed
insight into the fragmentation is also provided \citep{koopman2019}.
Therefore, GFN2-xTB is a promising candidate to study the
fragmentation of PAHs in an astronomical context.  The fragmentation
of PAHs is investigated through the following steps.
The first step is the preparation of a MD simulation.
A structure optimization using GFN2-xTB is performed for the ion of the interest.
The second step is the equilibration of the investigated ion.
An optimized structure is equilibrated at
500K and followed by a constant particle number, simulation
volume, and total energy run, a so-called NVE run. This
temperature is chosen to guarantee the adequate sampling of the
configuration space. A predefined number of snapshots
(i.e., 1000) during the NVE run are randomly selected for further
runs.
The third step is the excitation of ions.
For each selected snapshot, the excitation of ions is simulated by a simple "heating" in MD which brings the ion directly from the vibrational "cold" state to the vibrational "hot" state near the dissociation threshold. 
Specifically, the heating is imposed by applying a Berendsen thermostat \citep{berendsen1984} with the target temperature of 3000 K. This temperature is chosen so that a reasonable amount of fragments have been observed.
More discussions on the choice of the temperature can be found in Appendix. \ref{s-app-mass}, Fig. \ref{fig:more-md-mass-C16H15}, and Fig. \ref{fig:more-md-mass-C16H17}.
The heating time is set to 3 ps. The time constant in the Berendsen thermostat is set to 0.1 ps.
The final step is production runs in which hot ions are independently propagated for 10 ps on the ab initio potential energy surface (PES) in the NVE ensemble.
The number of fragments and the mass of each fragment are counted every 5 fs for a single run. Finally, the mass spectrum is generated by collecting the mass distribution of 1000 independent runs.
For the purposes of this study, the time step of all MD runs was set to 0.5 fs. Detailed discussions on simulation settings can be found in \citet{mass_theory_2013}.

Density functional theory (DFT) calculations were performed using
B3LYP \citep{becke1993a,lee1988} and the def2-TZVP basis set \citep{def2} within the quantum chemistry software ORCA 4.2.1 \citep{ORCA4}. 
The RIJCOSX approximation \citep{RIJCOSX} and def2/J auxiliary basis set \citep{def2-fit} were used to accelerate the Coulomb and HF exchange integration for the B3LYP calculation.
Dispersion interactions were corrected by the DFT-D3 method \citep{D3} with Becke-Johnson damping \citep{D3BJ}.
Geometry optimizations were performed with the standard convergence criteria ("Opt" keyword in ORCA).
The transition states were located through a combination of climbing image nudged elastic band (CI-NEB) \citep{CI-NEB} and eigenvector-following (EF) methods \citep{asgeirsson2021} implemented in ORCA.
All IR calculations were performed within the harmonic approximation.
A scaling factor of 0.9671 \citep{freq_scaling} was applied to the vibrational frequencies as an empirical way to include anharmonic effects.
Each vibrational peak was convoluted with a Gaussian line shape function with a full-width at half-maximum (FWHM) of 20 cm$^{-1}$.

\section{Results}
\subsection{Mass loss of 28} \label{s-mass}
\begin{figure}
        \centering
        \includegraphics[width=1\linewidth]{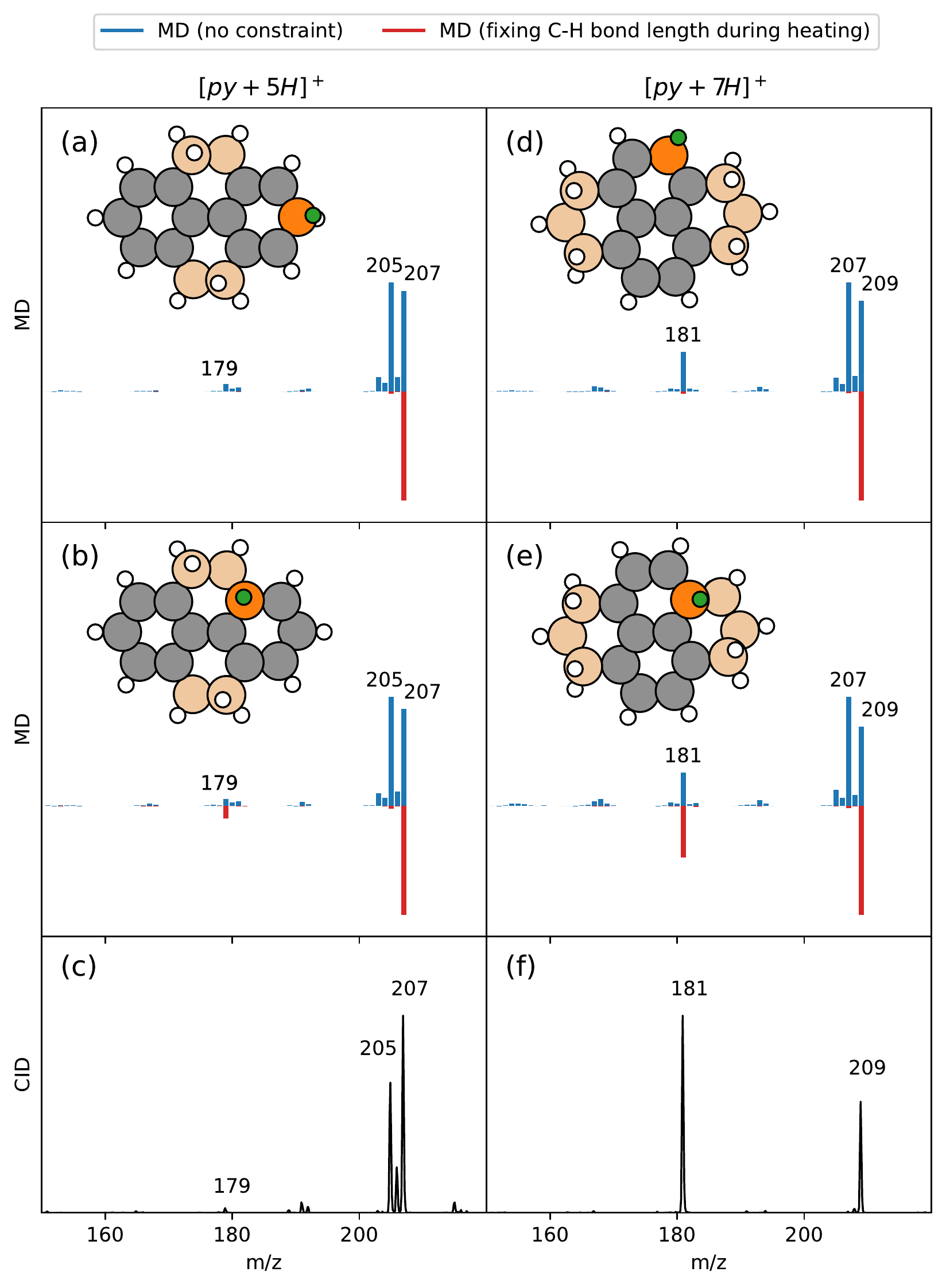}
        \caption{
        Comparisons of theoretical and experimental mass spectra of two protonated pyrenes.
        All theoretical mass spectra were extracted from MD simulations at
        (a) and (b): Two theoretical mass spectra from $[py+5H]^+$ with different protonated sites.
        The isomer in (a) is 0.28 eV more stable than the isomer in (b).
        The color scheme of carbon and hydrogen atoms are as follows: gray (aromatic carbon), light yellow (aliphatic carbon), orange (protonated carbon), white (normal hydrogen), and green (proton).
        (c): The experimental mass spectrum of $[py+5H]^+$recorded by collision-induced dissociation (CID).
        (d) and (e): Two theoretical mass spectra from $[py+7H]^+$ with different protonated sites.
        The isomer in (d) is 0.08 eV more stable than the isomer in (e).
        (f): The experimental mass spectrum of $[py+7H]^+$ recorded by CID.
        See Fig. \ref{fig:more-exp-mass} for the log-scaled mass spectra of (c) and (f)
        where peaks with lower intensities can be discerned.
    }
        \label{fig:AIMD-mass}
\end{figure}

The two superhydrogenated pyrene cations, $[py+5H]^+$ and $[py+7H]^+$,
exhibit significantly different fragmentation patterns, as shown in
the experimental mass spectra in Fig. \ref{fig:AIMD-mass} (c) and (f)
which were recorded by CID.
We note that $[py+5H]^+$ primarily shows the loss of two hydrogen atoms which is attributed to molecular hydrogen formation \citep{FELIX-H2}, while $[py+7H]^+$
only has one well-characterized mass loss of 28, which we expect to be
ethylene (C$_2$H$_4$).
To verify this hypothesis and probe the dissociation dynamics directly, MD with GFN2-xTB calculations were performed to generate theoretical mass spectra of two protonated pyrenes given in the first two rows of Fig. \ref{fig:AIMD-mass}.
For each protonated parent species, two protonated sites have been considered with the more stable isomers in Fig. \ref{fig:AIMD-mass} (a) and (d), as well as with the less stable isomers in Fig. \ref{fig:AIMD-mass} (b) and (e).
This was done since the exact structure of the protonated parent species is unknown and different protonated structures might contribute to the experimental mass spectra at the same time.
Initially, MD calculations were performed in a way without geometry constraints as described in Sect. \ref{s-Methods} so that the resulting fragmentation pattern would not be biased by our MD settings.
The obtained theoretical mass spectra are plotted in blue in Fig. \ref{fig:AIMD-mass}.
Both theoretical mass spectra of the two $[py+5H]^+$ isomers agree well with the experiment.
For $[py+7H]^+$, the dominant fragment in the experiment, m/z=181, has been reproduced but with a lower yield and the two isomers show no differences in their mass spectra.
However, a serious problem for the MD calculations for $[py+7H]^+$ is the high yield of mass peak 207 which originated from the 2H/H$_2$ loss of its parent ion.
This is in contradiction with the experimental observation which does not show 2H/H$_2$ loss at all.

One possible explanation for the observed 2H/H$_2$ loss in our GFN2-xTB based MD
simulations may be a too facile C-H bond breaking at this level of
theory. The facile C-H bond breaking has been observed in another empirical
potential OM2 when electron impact mass spectra (EI-MS) were simulated \citep{mass_theory_2013}.
To test this conjecture and to investigate the implications of
having less facile C-H bond breaking, we performed a set of GFN2-xTB based
MD simulations in which we constrained the C-H bonds to fixed
nonbreakable bond lengths during the heating phase of MD
simulations.
We find that this procedure changes the observed behavior in the subsequent production phase of the run.
Here, the C-H bonds are allowed to change, 
but due to the better equilibration during the heating phase,
the C-H bonds now are less prone to break during the production run where the mass of each fragment at each MD step is counted.

The mass spectra resulting from this "gedanken experiment" are shown
in red in Fig. \ref{fig:AIMD-mass} with a flipped y axis.
The fragment masses corresponding to hydrogen losses indeed disappear
for both $[py+5H]^+$ and $[py+7H]^+$ for such simulations having
constrained C-H bonds in the heating phase. We shall not pursue this
issue further in the present work, but rather focus on the fragments other
than the 2H/H$_2$ related ones that originate from either type of
simulation.
Since the topic of this paper is ethylene formation, we focus our attention on the fragment with m/z=179 for $[py+5H]^+$ and the fragment with m/z=181 for $[py+7H]^+$. We note that
$[py+7H]^+$ is more efficient at producing a mass loss of 28 than $[py+5H]^+$, as can be seen when comparing Fig. \ref{fig:AIMD-mass} (b) and (e).

\subsection{Confirmation of ethylene formation} \label{s-C2H4}
\begin{figure}
        \centering
        \includegraphics[width=1\linewidth]{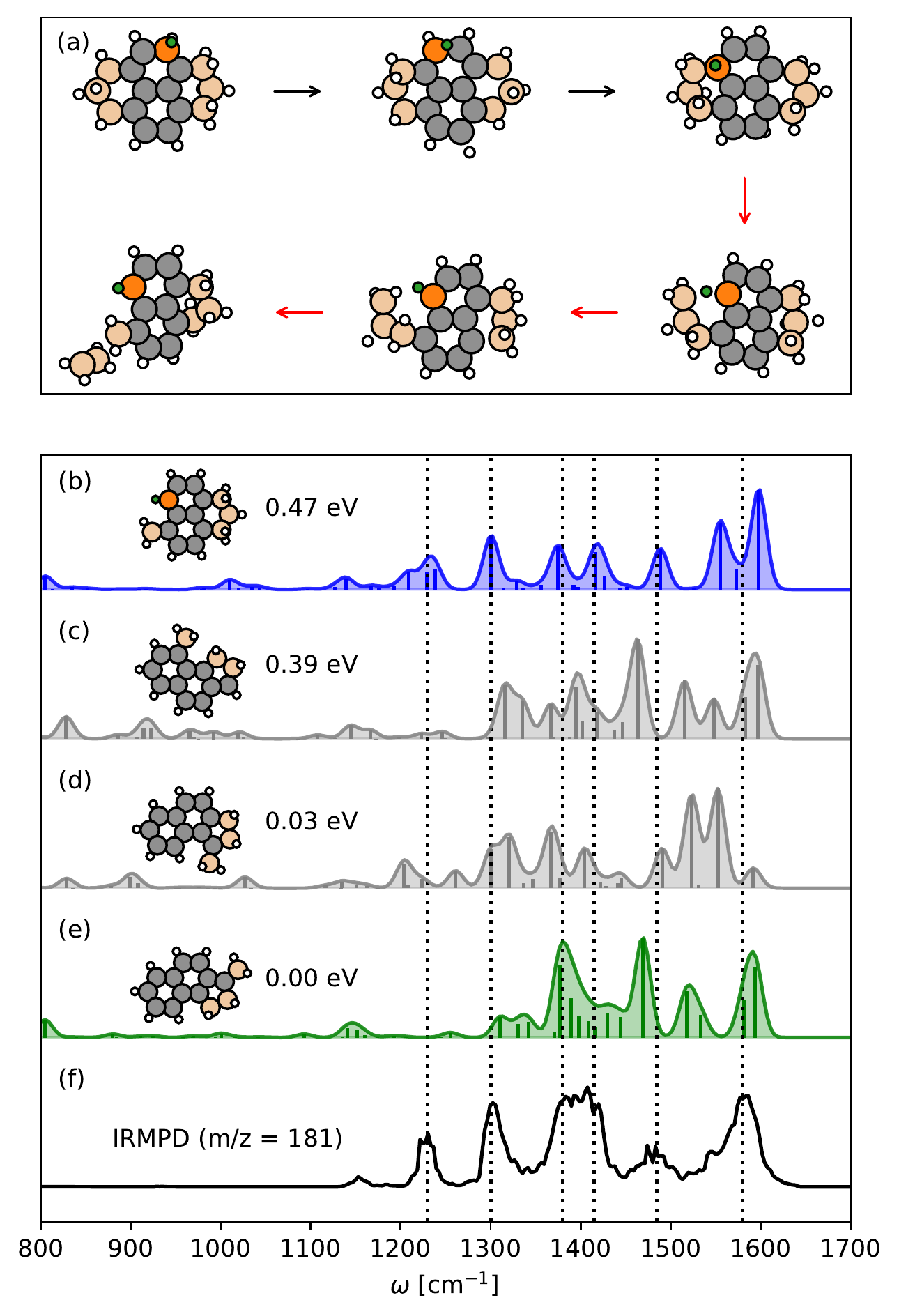}
        \caption{
        Identification of the fragment (m/z=181).
        (a) The fragmentation pattern of $[py+7H]^+$ extracted from MD trajectories.
        The color scheme for atoms is the same as the scheme in Fig. \ref{fig:AIMD-mass}.
        The hydrogen diffusion is indicated by black arrows and the process of ethylene formation is indicated by red arrows.
        The computed IR spectrum of (b) the MD-determined isomer, (c) the EA-determined isomer, (d) the EA-determined isomer, and (e) the EA-determined isomer with their structure snapshots and relative energies.
        (f) The experimental IR spectrum of the fragment (m/z=181).
        Vertical dashed lines have been drawn to aid the comparison between theory and experiment.
    }
        \label{fig:IR-frag-181}
\end{figure}

Ethylene formation is the only possible channel for mass loss of 28 in $[py+7H]^+$ with and without geometry constraints during heating as extracted from the MD trajectories in Fig. \ref{fig:IR-frag-181} (a).
The added proton easily jumps from one carbon atom to another in the molecule, and ethylene is only ejected when the proton is bonded to the carbon next to the desorbed C$_2$H$_4$ group.
The same hydrogen migration is thought to contribute to H$_2$ loss \citep{chen2019} and CH$_n^+$(n=4-6) fragments \citep{chacko2020} from PAHs.
Further confirmation of ethylene formation is achieved through the IR calculations of the fragment (m/z=181) which are shown in Fig. \ref{fig:IR-frag-181} (b)-(f).
To explore the configuration space of the fragment (m/z=181), a global optimization method, the evolutionary algorithm (EA) \citep{vilhelmsen2014a,jorgensen2017}, was used to generate as many candidates as possible.
Selected candidates from the EA approach and one candidate from the MD runs corresponding to the ethylene formation were subsequently screened by harmonic IR calculations.
The IR features of the candidate from the MD search (blue color, Fig. \ref{fig:IR-frag-181} (b)) have better agreement with the experiment (Fig. \ref{fig:IR-frag-181} (f))
than the most stable configuration (global minimum, plotted in green, Fig. \ref{fig:IR-frag-181} (e)) and other candidates (gray color, Fig. \ref{fig:IR-frag-181} (c) and Fig. \ref{fig:IR-frag-181} (d)) found by the EA search.
The band assignments of the MD candidate were achieved by visualizing its normal modes (see Fig. \ref{fig:more-IR-frag-181}). 
The comparison between the IR spectrum of the parent ion $[py+7H]^+$ and that of the fragment (m/z=181) is shown in Fig. \ref{fig:more-IR}.
A computed band of the MD candidate at 1140 cm$^{-1}$ is due to the in-plane bending of aromatic C-H.
A series of bands in the 1200 - 1500 cm $^{-1}$ region have contributions from C-C stretching, C-H in-plane bending ,and CH$_2$ bending.
The computed band positions at 1229, 1301, 1375, 1417, and 1488 cm$^{-1}$ in Fig. \ref{fig:IR-frag-181} (b) agree well with the IRMPD peaks at 1230, 1300, 1380, 1415, and 1485 cm$^{-1}$ in Fig. \ref{fig:IR-frag-181} (f).
The bands above 1500 cm$^{-1}$ involve aromatic C-C stretching and aromatic C-H in-plane bending.
The computed peak at 1573 cm$^{-1}$ is the closet one to the 1580 cm$^{-1}$ IRMPD band, even though its relative band intensities differ a lot.
For the other three isomers, their bands may fit well with some of the IRMPD bands.
However, they either have bands that are absent in the IRMPD spectrum or fail to reproduce some IRMPD peaks.
Given that the agreement between Fig. \ref{fig:IR-frag-181} (b) and Fig. \ref{fig:IR-frag-181} (f) is already good,
we refrained from constructing a blended IR spectrum composed of contributions from numerous isomers.
Such an attempt was carried out by \citet{wenzel2022} in another H-functionalized pyrene system and was found challenging as it requires choices to be made about the number of isomers to include and for their relative weights.
Based on the good match between Fig. \ref{fig:IR-frag-181} (b) and Fig. \ref{fig:IR-frag-181} (f), we use the MD-determined isomer (Fig. \ref{fig:IR-frag-181} (b)) in the reminder of the paper.
It has 0.47 eV higher energy than the global minimum.
This implies that the fragment (m/z=181) might stay in a local minimum region without opportunities or enough energy to jump into the global minimum after losing ethylene from $[py+7H]^+$.
Combining the results from MD and IR calculations, it is clear that ethylene is responsible for the mass loss of 28 for $[py+7H]^+$.
The same condition for ethylene formation is valid for $[py+5H]^+$, but the difference is that the yield of ethylene is much lower than for $[py+7H]^+$, as can be observed from the extremely low intensity of the fragment (m/z=179) in Fig. \ref{fig:AIMD-mass} (a) and (b).

\subsection{Role of the edge structure} \label{s-edge} 
\begin{figure*}
        \centering
        \includegraphics[width=1\textwidth]{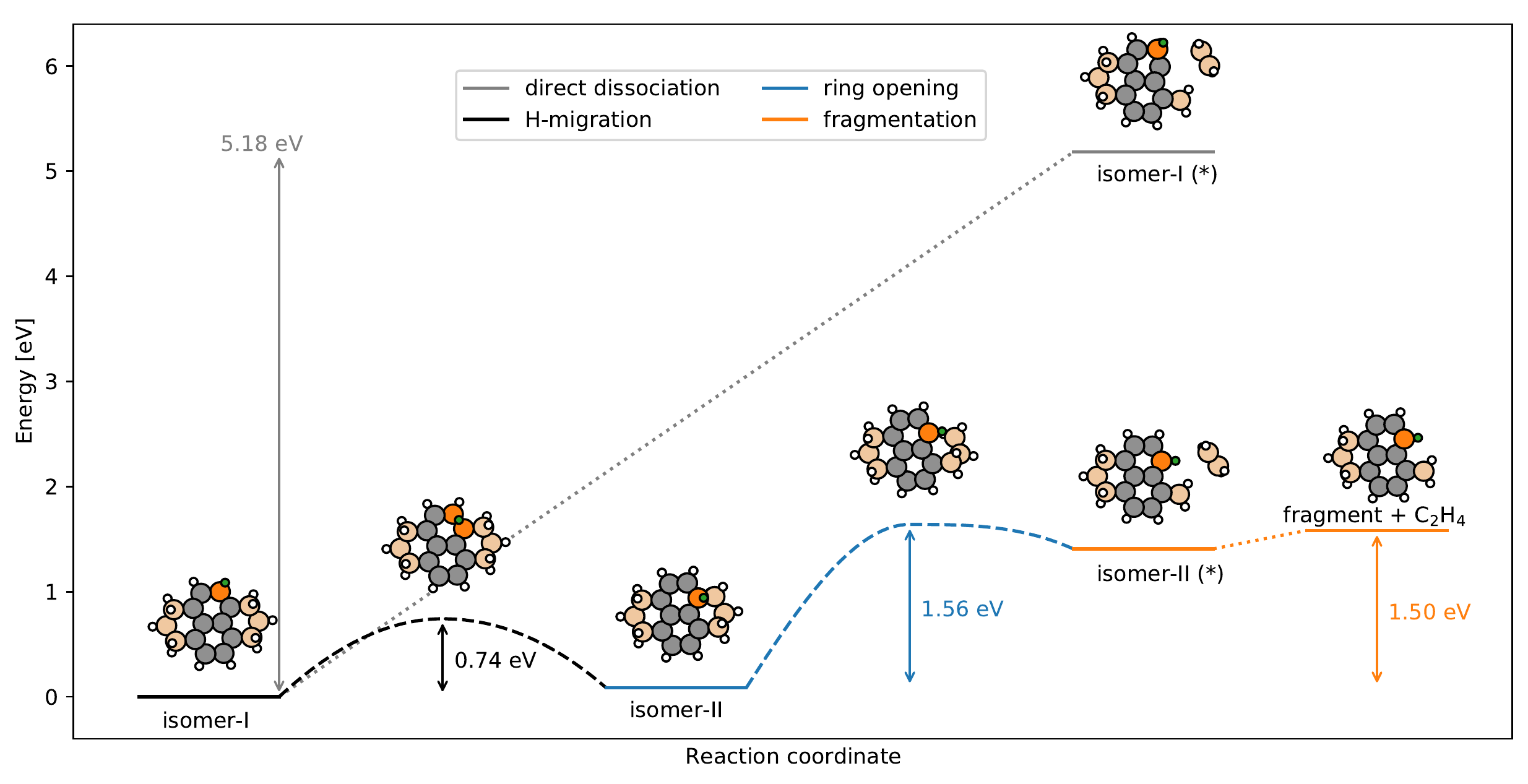}
    \caption{
      Energy profile along the pathway of C$_2$H$_4$ formation from $[py+7H]^+$.
      Each horizontal line corresponds to a local minimum.
      If a curved line is plotted between two horizontal lines, it represents a transition path with the energy barrier indicated underneath.
      The color scheme of atoms is the same as that of Fig. \ref{fig:AIMD-mass}.
      }
        \label{fig:barrier-C16H17-proton}
\end{figure*}

As observed in
MD calculations, ethylene formation is heavily dependent on the
geometry, especially the edge structure (position of hydrogenated and
protonated sites), of these superhydrogenated PAHs. We note that  $[py+7H]^+$ has
superhydrogenated carbon in trio groups (three connected hydrogenated
carbon and associated hydrogen atoms are referred to as a trio group).
Ethylene is efficiently produced when the proton migrates to the
carbon next to the trio group (Fig. \ref{fig:IR-frag-181} (a) and
isomer-II in Fig. \ref{fig:barrier-C16H17-proton}).  To understand
this process, energy barrier calculations using DFT were performed.
The most stable isomer of $[py+7H]^+$ is isomer-I in
Fig. \ref{fig:barrier-C16H17-proton}, which has protonated carbon
that is not directly connected with the trio group.  Isomer-I needs at least
5.18 eV to produce ethylene directly.  Such a large energy barrier is due to
the high-energy structure of the dissociated product which has broken
aromaticity due to the unfavorable proton position and unsaturated
carbon.  Alternatively, isomer-I can also undergo a hydrogen migration
before the ethylene formation.  After overcoming an energy barrier of
0.74 eV, the hydrogen migration leads to the formation of isomer-II,
which requires only 1.56 eV to open the ring and expel ethylene.
The role of hydrogen migration has also been addressed in the ring opening and 
the dissociation of a H-functionalized PAH cation by MD simulations \citep{jusko2018}.
The fragmentation energy (E$_\text{fragment}$ + E$_{\text{C}_2\text{H}_4}$ - E$_\text{parent ion}$) of isomer-II 
is 1.50 eV, as shown in Fig. \ref{fig:barrier-C16H17-proton}. 
Therefore, the requirement of ethylene formation is 
that the protonated carbon atom be next to the trio group.  
The importance of neighboring hydrogen to the trio group and hydrogen
migration is also observed for 1, 2, 3, 4-tetrahydronaphthalene \citep{diedhiou2019},
which needs to overcome a 1.06 eV energy barrier. 

\begin{figure*}
        \centering
        \includegraphics[width=1\textwidth]{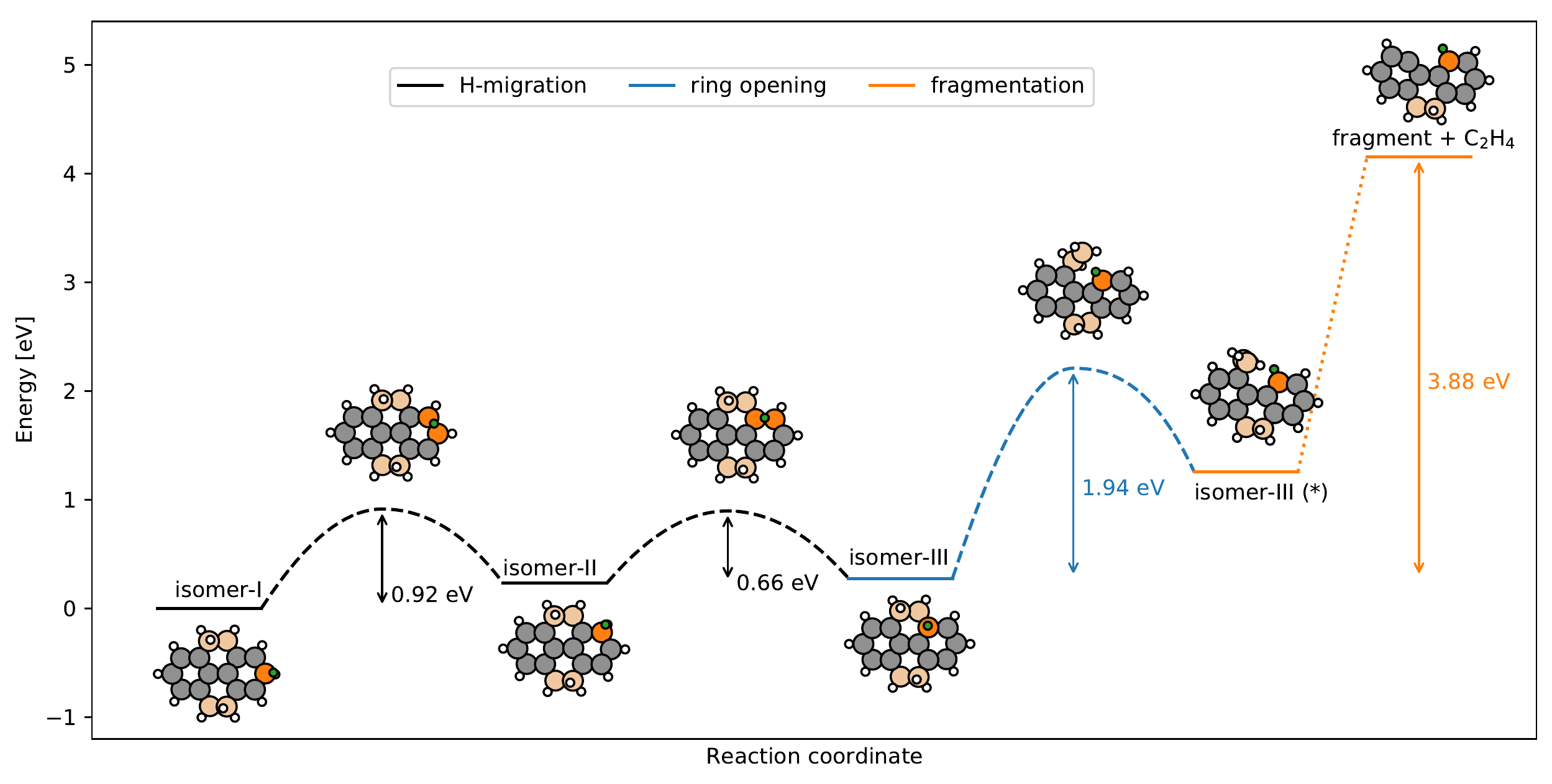}
    \caption{
      Energy profile along the pathway of C$_2$H$_4$ formation from $[py+5H]^+$.
      Caption is similar to that of Fig. \ref{fig:barrier-C16H17-proton}.
      }
        \label{fig:barrier-C16H15-proton}
\end{figure*}

The pathway of C$_2$H$_4$ formation from $[py+5H]^+$ is given in Fig. \ref{fig:barrier-C16H15-proton}. We note that
$[py+5H]^+$ can also follow similar hydrogen migration as $[py+7H]^+$, 
but it has superhydrogenated carbon in duo groups (two connected hydrogenated carbon and associated
hydrogen atoms are referred to as a duo group) before C$_2$H$_4$ formation.
The edge structure becomes slightly different from that of $[py+7H]^+$.  
The energy barrier for opening the ring is 1.94 eV, 
but the fragmentation energy reaches 3.88 eV.  
In contrast to the dissociation
from $[py+7H]^+$, C$_2$H$_4$ is still covalently attached to the carbon skeleton
in $[py+5H]^+$ after the ring is opened.  Even if one C-C bond
breaks up, the huge fragmentation energy still renders the ethylene
detachment unlikely.  
The unsaturated carbon atom of the fragment after C$_2$H$_4$ formation 
is the possible cause of such a high fragmentation energy.
The high fragmentation energy agrees well with the extremely low yield of C$_2$H$_4$ loss of $[py+5H]^+$ in Fig. \ref{fig:AIMD-mass}.
This implies that the CID experiments cannot provide enough energy to detach C$_2$H$_4$ from $[py+5H]^+$,
while the C$_2$H$_4$ formation from $[py+7H]^+$ with a computed fragmentation energy of 1.56 eV is feasible in the experiment setup.
Hence, the edge structure of superhydrogenated pyrene is an important
factor in determining ethylene formation.

\subsection{Other superhydrogenated PAHs} \label{s-PAHs}
\begin{figure}
        \centering
        \includegraphics[width=1\linewidth]{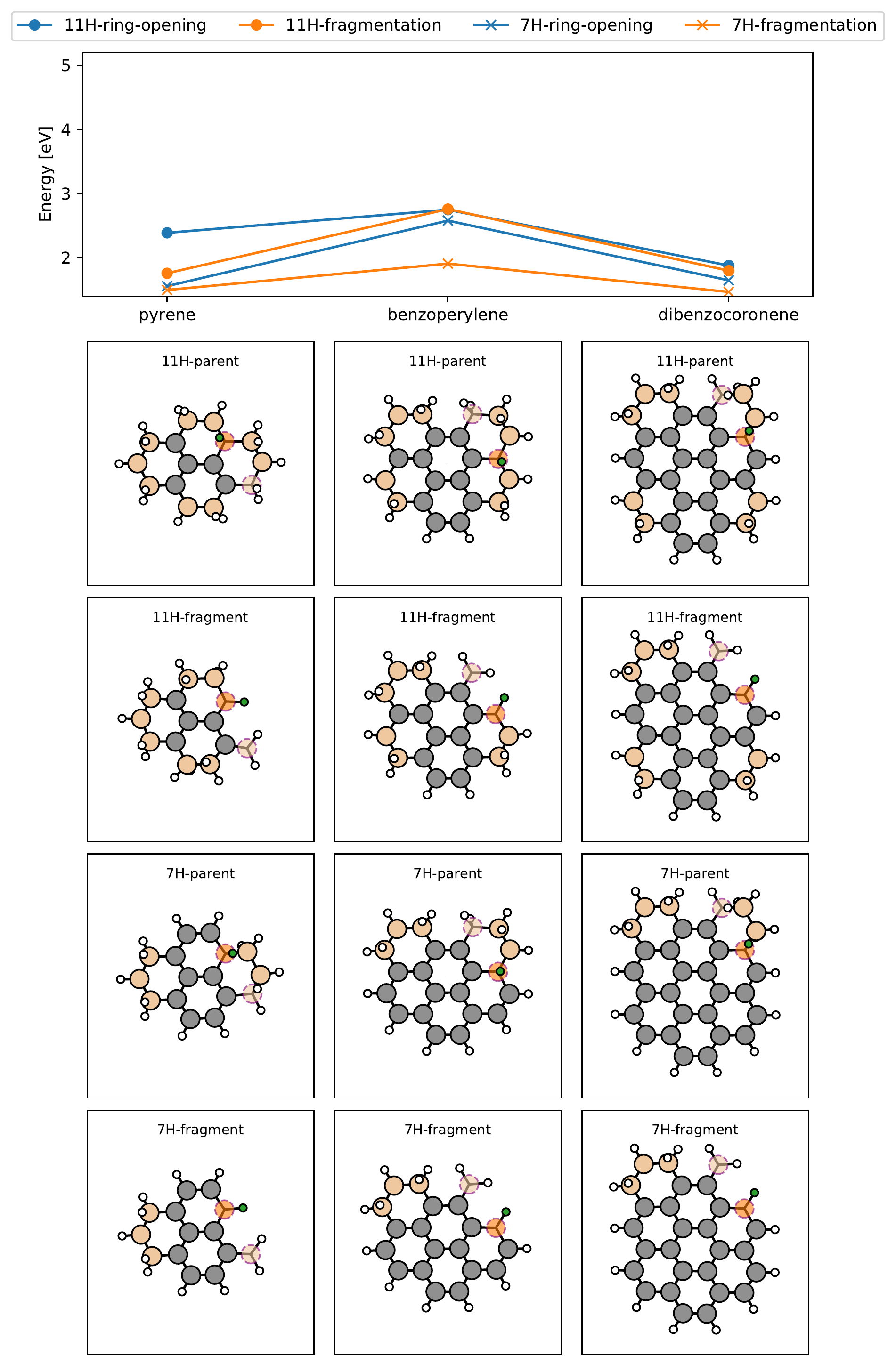}
        \caption{
    Energy barriers of opening a ring and fragmentation energies of C$_2$H$_4$ formation on three hydrogenated PAHs (pyrene, benzoperylene, and dibenzocoronene) with the hydrogenated state in trio groups.
    Structure snapshots of parent ions and fragments are plotted in the bottom panel.
    The connecting carbon atoms to C$_2$H$_4$ are highlighted by purple edges and more transparent colors than normal atoms.
        }
        \label{fig:barrier-7H}
\end{figure}

\begin{figure}
        \centering
        \includegraphics[width=1\linewidth]{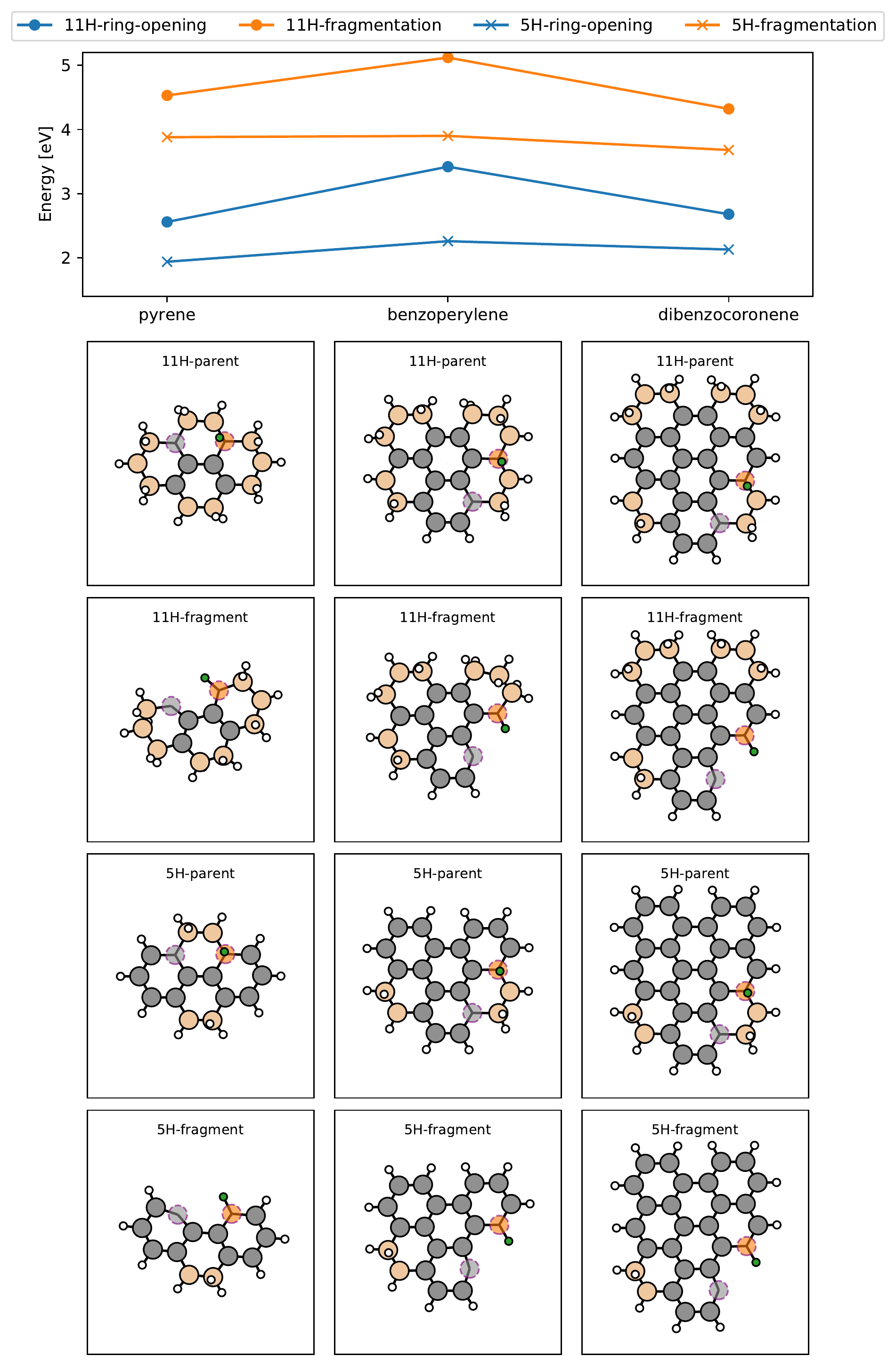}
        \caption{
        Energy barriers of opening a ring and fragmentation energies of C$_2$H$_4$ formation on three hydrogenated PAHs (pyrene, benzoperylene, and dibenzocoronene) with the hydrogenated state in duo groups.
        Structure snapshots of parent ions and fragments are plotted in the bottom panel.
        The connecting carbon atoms to C$_2$H$_4$ are highlighted by purple edges and more transparent colors than normal atoms.
        }
        \label{fig:barrier-5H}
\end{figure}

As an attempt to reveal the general rules for ethylene formation from superhydrogenated PAHs, energy barrier calculations were performed on two more hydrogenated PAHs which are built on top of the pyrene structure.
These two hydrogenated PAHs (benzoperylene and dibenzocoronene) were chosen because they have edge structures both in duo and trio groups and they vary in molecular size. 
Fig. \ref{fig:barrier-7H} shows the energy barriers of opening a ring and fragmentation energies of C$_2$H$_4$ formation for three hydrogenated PAHs ($[py+7H]^+$, $[benzo+7H]^+$, and $[dibenzo+7H]^+$) with the hydrogenated state in trio groups.
The C$_2$H$_4$ fragmentation energy remains almost unchanged as the size of the PAHs increases.
In the case of a duo group (shown in Fig. \ref{fig:barrier-5H}), all PAHs have C$_2$H$_4$ fragmentation energies above 3.6 eV, which is much higher than the level of PAHs in the trio group (less than 2 eV).
Such high C$_2$H$_4$ fragmentation energies are caused by unsaturated carbon atoms after C$_2$H$_4$ is released from PAHs.

Further hydrogenation up to 11 additional hydrogen atoms in total has allowed a direct comparison between trio and duo groups in ejecting C$_2$H$_4$ from $[py+7H]^+$.
The loss of ethylene is still easier from a trio than from a duo group.
Among all three PAHs studied here, the energy barrier climbs as the degree of hydrogenation becomes higher, meaning that the more hydrogenated the edge is, the more difficult it is for the reaction to occur.
In addition, C$_2$H$_4$ fragmentation energies almost remain unchanged when varying the degree of hydrogenation.

\section{Astrophysical implications} \label{s-astro}
Our calculations and experiments reveal a top-down formation process of ethylene from the hydrogenated edge of PAHs.
This process has a very strict geometry requirement which is the hydrogenated edge as a trio group with a neighboring proton.
The improved activity in producing ethylene from superhydrogenated pyrene is achieved by increasing the number of additional hydrogen atoms from 5 to 7.
This is not due to the higher degree of hydrogenation, but it is associated to the geometry requirement only met by $[py+7H]^+$.
It is still under debate whether the hydrogenation can protect PAHs under the harsh interstellar environment.
A study of soft x-ray absorption in pristine and hydrogenated coronene cations, C$_{24}$H$_{12+m}^+$ (m = 0–7), led to the conclusion that additional hydrogen atoms protect the carbon backbone from fragmentation \citep{reitsma2014}.
Another study conducted collision experiments between fast (30–200 eV) He atoms and pyrene (C$_{16}$H$_{10+m}^+$ m = 0, 6, and 16) and proposed an opposite conclusion that the increased degree of hydrogenation led to more carbon backbone fragmentation \citep{gatchell2015}.
Based on the requirement for ethylene formation, the hydrogenated coronene has no way to generate hydrogenated edges in trio, thus making it unlikely to break the carbon backbone.
The superhydrogenated pyrene is different because of the possibility of having edges with trio groups.
DFT calculations \citep{gatchell2015} confirmed that C$_2$H$_4$ loss is more favorable than H loss from C$_{16}$H$_{16}^+$, even though a neighboring proton is not present.
The contrary conclusions from studies of superhydrogenated pyrene and coronene might be attributed to the different edge structure of pyrene and coronene.
It is not enough to draw a general conclusion just based on one or two PAHs.
More investigations are needed to understand the protection mechanism for PAH molecules.

The detection of C$_2$H$_4$ was first reported by \citet{betz1981} toward IRC+10216.
A mid-IR observational study of C$_2$H$_4$ toward IRC+10216 by \citet{fonfria2017} revealed that the C$_2$H$_4$ abundance relative to H$_2$ is between $10^{-8}$ and $10^{-7}$.
From the view of top-down chemistry, H$_2$ formation from PAHs does not deplete carbon atoms in PAHs so that PAHs can be repeatedly used to generate H$_2$. The loss of
C$_2$H$_4$  from PAHs must take away carbon atoms from PAHs, thus following C$_2$H$_4$ formation is prohibited due to the depletion of carbon atoms.
In addition, C$_2$H$_4$ formation is more sensitive to hydrogenated sites of PAHs than H$_2$ formation.
It is expected that H$_2$ formation has more viable routes than C$_2$H$_4$ formation via top-down chemistry of PAHs.

\section{Conclusions} \label{s-conclusion}
It has been shown in this work that MD combined with GFN2-xTB is an efficient method to produce qualitatively correct mass spectra.
The dissociation mechanisms have been accessed by directly extracting MD trajectories and verified by IR calculations and experiments on fragments.
An advantage of this MD approach is that it is capable of probing dissociation events frame by frame.
The fragment comes out in a more natural way than global minimum optimization. 
MD simulations have shown that the C$_2$H$_4$ loss is strongly dependent on the geometry of superhydrogenated pyrenes.
Only superhydrogenated pyrenes which have the proton located on the carbon next to the trio group give a significant C$_2$H$_4$ yield.
Further DFT calculations lead to the conclusion that this geometry requirement is also valid for ethylene formation from other superhydrogenated PAHs.
This work demonstrates the importance of the hydrogenated edges in the production of small hydrocarbons from the processing of superhydrogenated PAHs.
Including this pathway in chemical models may reduce the disagreement between chemical models and observations in estimating the C$_2$H$_4$ abundance \citep{fonfria2017}.
In relation to PAH evolution in space, this study provides more evidence on the weakening of the carbon backbone due to the presence of aliphatic C-H bonds which are also reported in superhydrogenated PAHs \citep{gatchell2015} or alkylated PAHs \citep{marciniak2021}.
\begin{acknowledgements}
    The work is supported by the Danish National Research Foundation through the Center of Excellence “InterCat” (Grant agreement no.: DNRF150) and the European Union (EU) and Horizon 2020 funding award under the Marie Skłodowska-Curie action to the EUROPAH consortium, grant number 722346.
    We acknowledge support from VILLUM FONDEN (Investigator grant, Project No.\ 16562).
\end{acknowledgements}

\bibliography{ref}
\bibliographystyle{aa}

\begin{appendix}
  
\section{Mass spectra}\label{s-app-mass}
\begin{figure}[hbp]
    \centering
    \includegraphics[width=1\linewidth]{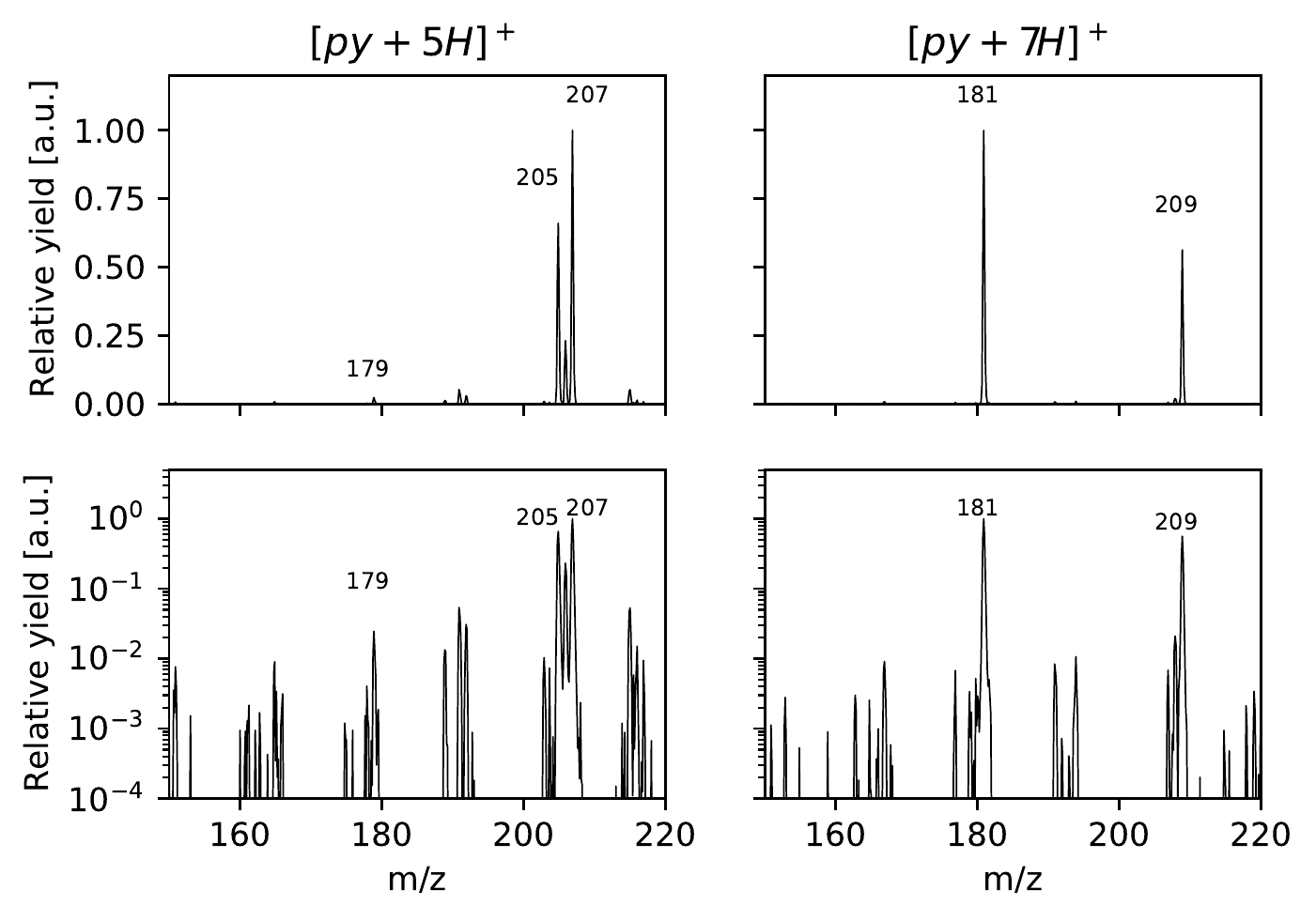}
    \caption{
          Experimental mass spectra of $[py+5H]^+$ and $[py+7H]^+$ in linear and logged scales.
      }
    \label{fig:more-exp-mass}
  \end{figure}

\begin{figure}
    \centering
    \includegraphics[width=1\linewidth]{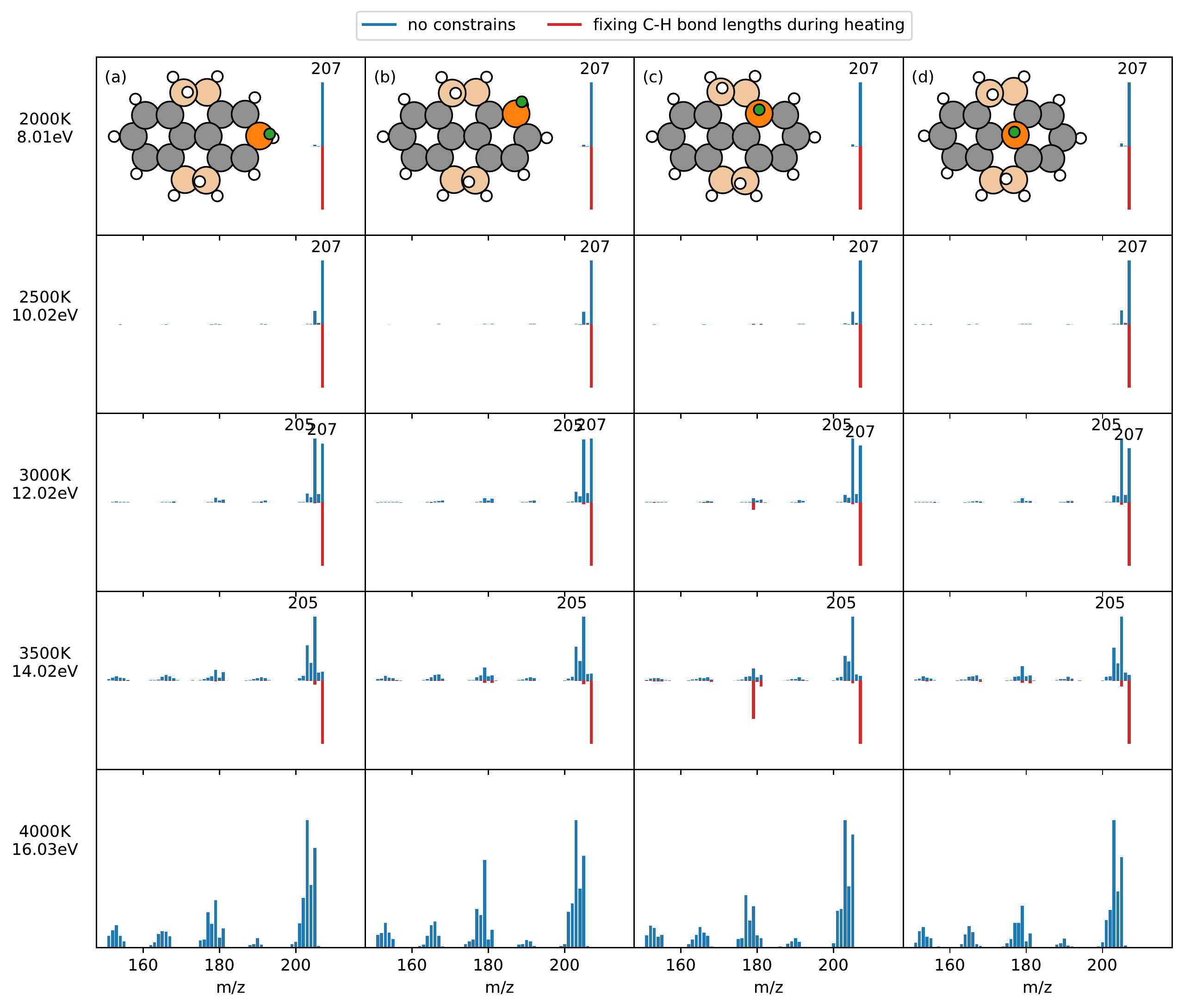}
    \caption{
          Theoretical mass spectra of different isomers of $[py+5H]^+$ at 2000, 2500, 3000, 3500, and 4000K.
          Temperatures and corresponding kinetic energies for each isomer are labeled on the y axis.
          The structures of the investigated isomers are shown in the upper panel.
      }
    \label{fig:more-md-mass-C16H15}
  \end{figure}

\begin{figure}
    \centering
    \includegraphics[width=1\linewidth]{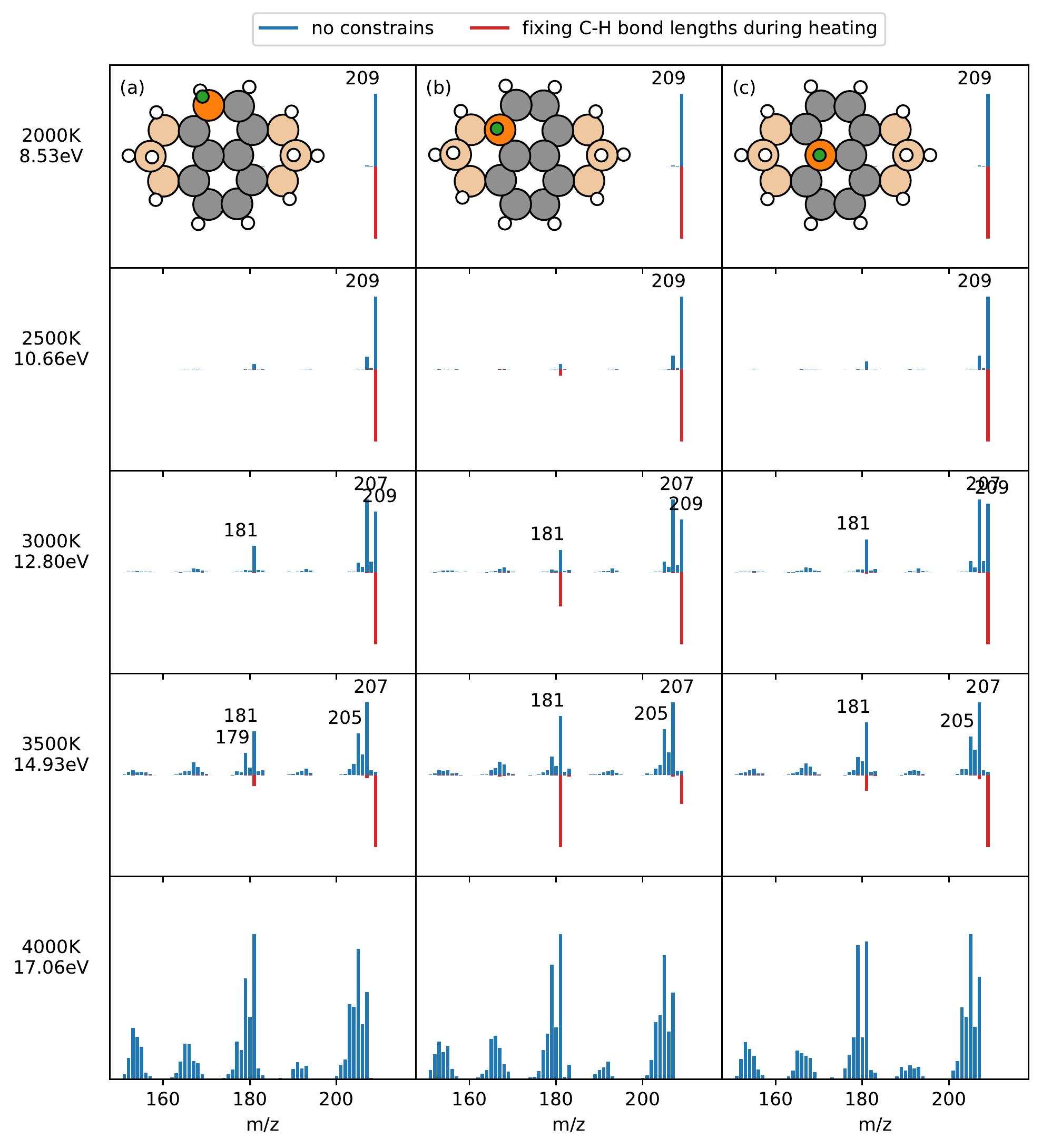}
    \caption{
      Theoretical mass spectra of different isomers of $[py+7H]^+$ at 2000, 2500, 3000, 3500, and 4000K.
      Temperatures and corresponding kinetic energies for each isomer are labeled on the y axis.
      The structures of the investigated isomers are shown in the upper panel.
      }
    \label{fig:more-md-mass-C16H17}
  \end{figure}

This section contains detailed mass spectra from experiments and theory.
Fig. \ref{fig:more-exp-mass} displays the experimental mass spectra of $[py+5H]^+$ and $[py+7H]^+$ in linear and logged scales.
The 2H/H$_2$ loss in $[py+5H]^+$ will be discussed in another study by \citet{FELIX-H2}.
The mass peaks relevant to C$_2$H$_4$ loss are highlighted.
Other mass peaks with low relative yields do appear in experiments and they are not discussed further.
Fig. \ref{fig:more-md-mass-C16H15} and Fig. \ref{fig:more-md-mass-C16H17} show the theoretical mass spectra of $[py+5H]^+$ and $[py+7H]^+$
using different protonated isomers and five selected temperatures (2000, 2500, 3000, 3500, and 4000K).
For both pyrene systems, the MD simulations at 2000K and 2500K fail to reproduce the relative yields of fragments,
while those at 3500K and 4000K tend to dissociate ions more than expected.
The 3000K MD simulation is the closest one to the CID experiment and was therefore shown in Fig. \ref{fig:AIMD-mass} and discussed in Sect. \ref{s-mass}.
For $[py+7H]^+$, different isomers do not show varied fragmentation patterns when there are no constrains.
When C-H bond lengths are fixed during heating phase, isomer b in Fig. \ref{fig:more-md-mass-C16H17} gives a higher yield of the fragment (m/z = 181) than other isomers.
This is due to the favorable protonated site of isomer b and is further explained in Sect. \ref{s-mass}.
\citet{FELIX-H2} will show how the choice of protonated isomers affects the fragmentation pattern of $[py+5H]^+$.
  
\clearpage
  \section{IR band assignments}
\begin{figure}[hbp]
    \centering
    \includegraphics[width=1\linewidth]{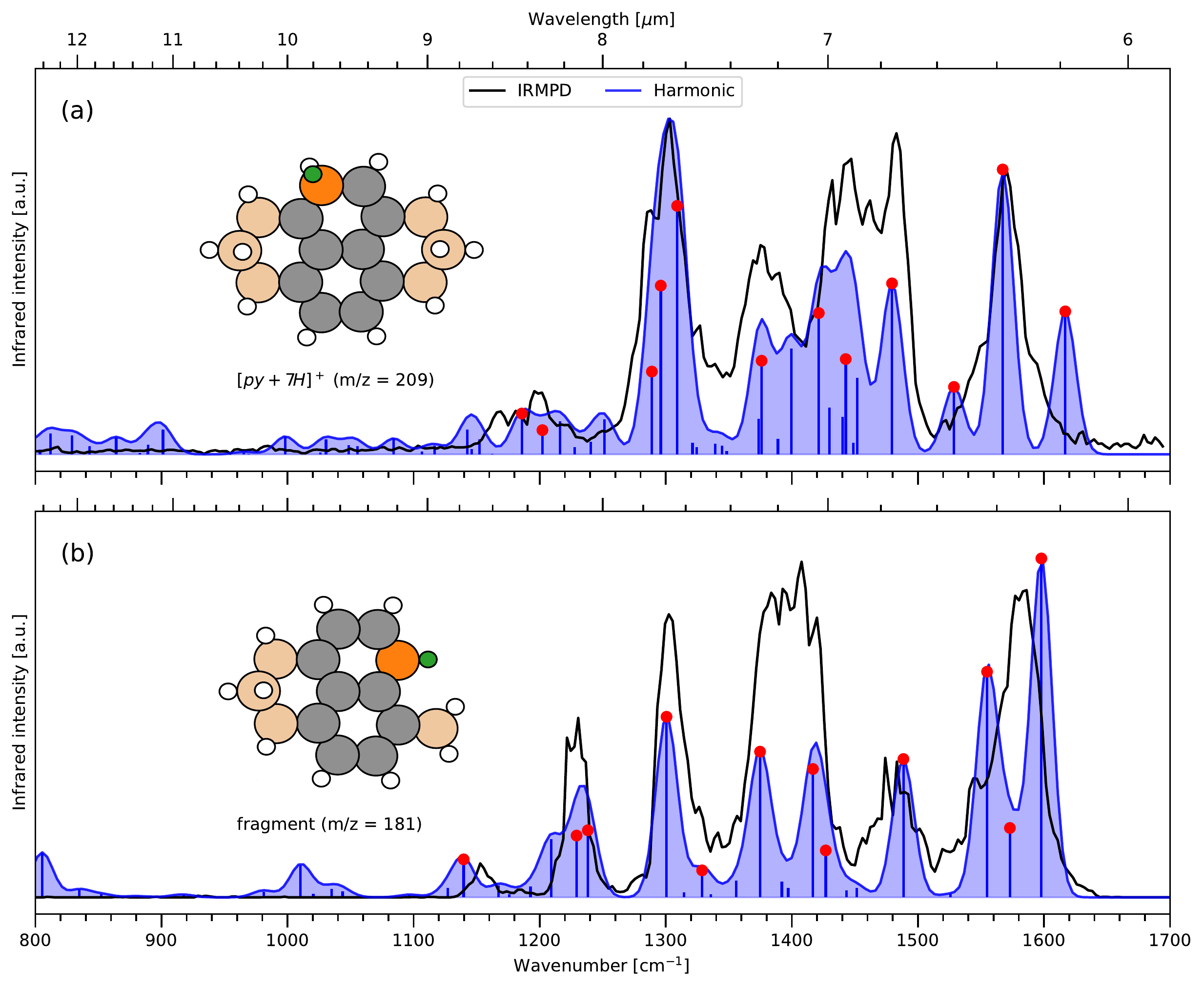}
    \caption{
          IR spectra of (a) $[py+7H]^+$ and (b) the fragment (m/z = 181).
          Visualizations of selected harmonic vibrational modes (red circles) are given in Fig. \ref{fig:more-IR-C16H17} and Fig. \ref{fig:more-IR-frag-181}.
      }
    \label{fig:more-IR}
  \end{figure}

\begin{figure*}
    \centering
    \includegraphics[width=1\textwidth]{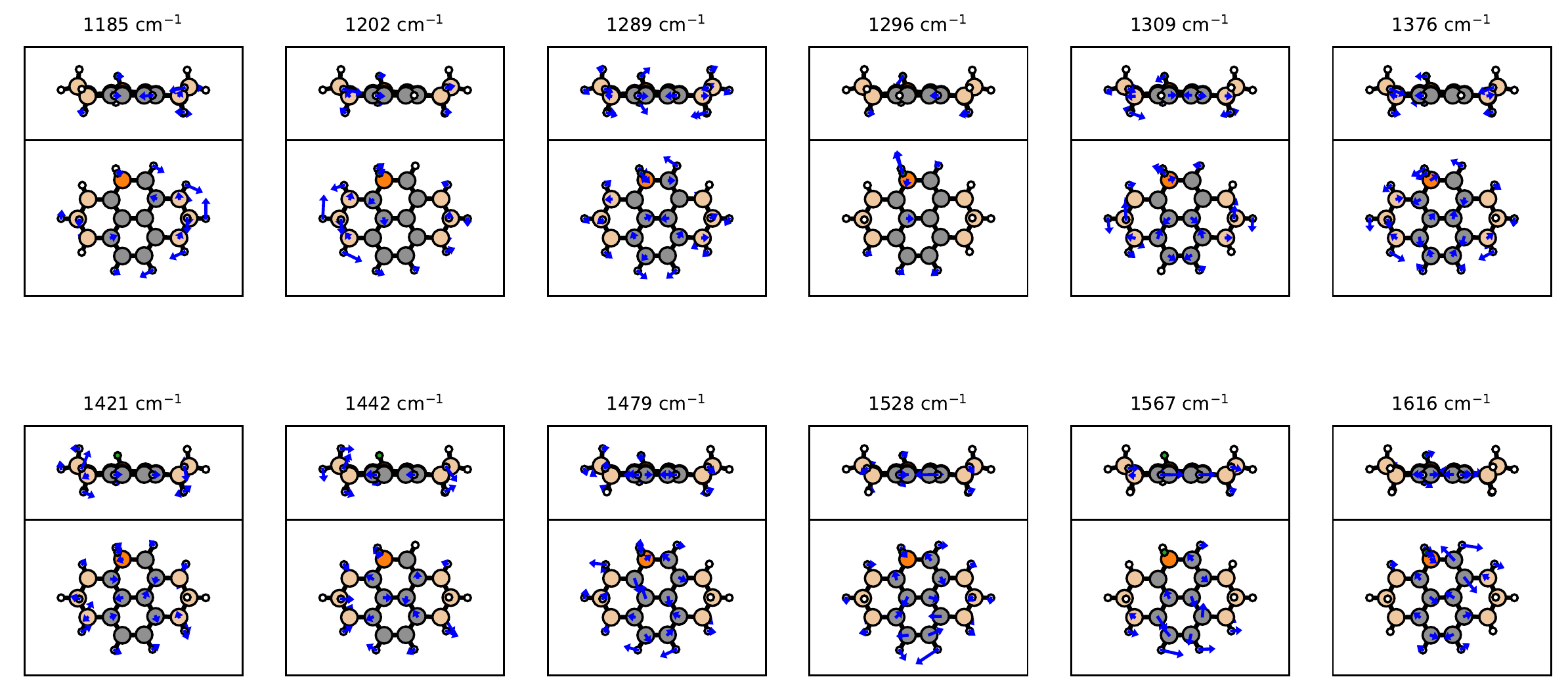}
    \caption{
          Selected harmonic vibrational modes of the most stable isomer of $[py+7H]^+$.
          Computed harmonic frequencies are labeled on top of the structure snapshots.
      }
    \label{fig:more-IR-C16H17}
  \end{figure*}

\begin{figure*}
    {
    \centering
    \includegraphics[width=1\textwidth]{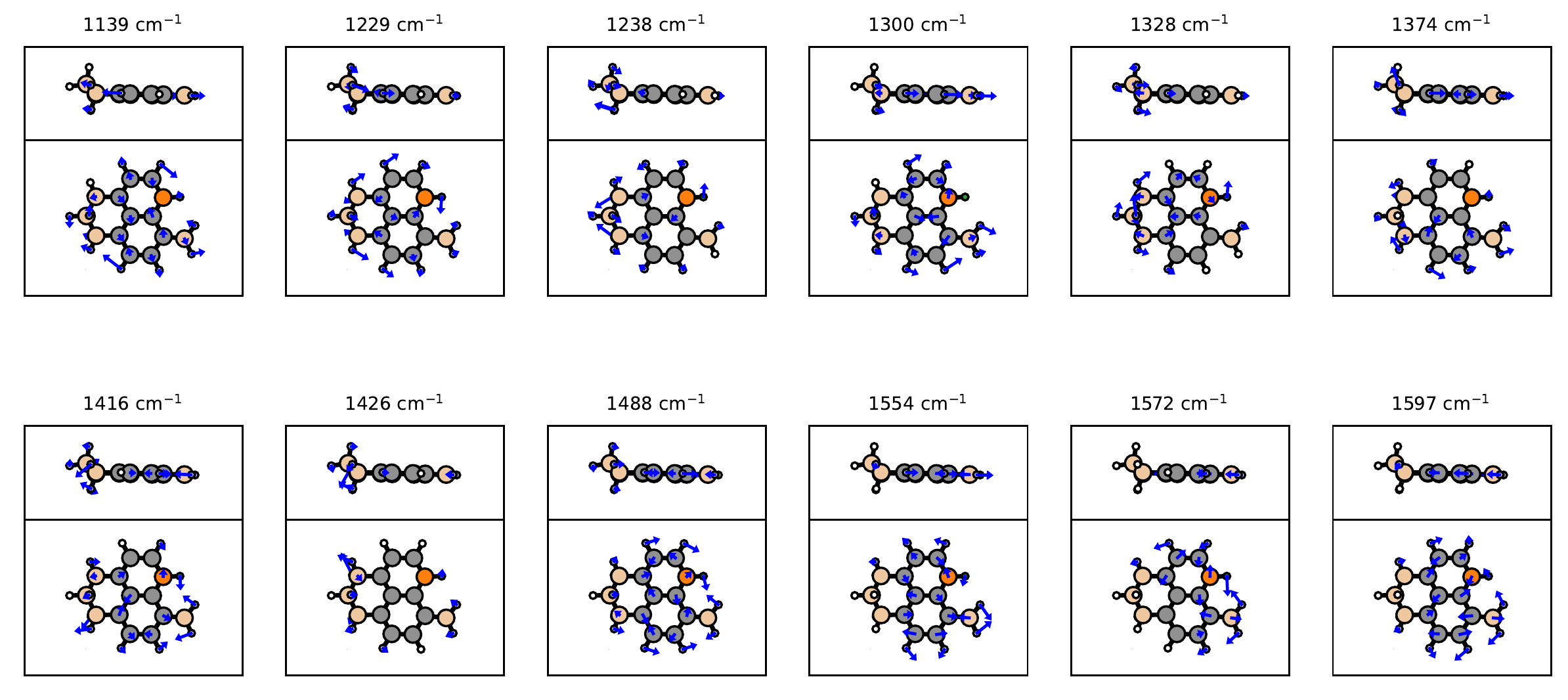}
          }
    \caption{
      Selected harmonic vibrational modes of the MD candidate of the fragment (m/z = 181).
      Computed harmonic frequencies are labeled on top of the structure snapshots.
      }
    \label{fig:more-IR-frag-181}
  \end{figure*}

This section compares IR spectra of $[py+7H]^+$ and (b) the fragment (m/z = 181) in Fig. \ref{fig:more-IR}
and includes their IR band assignments in Fig. \ref{fig:more-IR-C16H17} and Fig. \ref{fig:more-IR-frag-181}.
Both $[py+7H]^+$ and the fragment (m/z = 181) are rich in C-H in-plane bending and C-C stretching modes.
The main difference between them is where the CH$_2$ group apart from the trio group is located. We note that
$[py+7H]^+$ has the CH$_2$ group at the protonated site as a part of the ring, resulting in CH$_2$ scissoring mode at 1296 cm$^{-1}$
which is absent in the computed IR spectrum of the fragment (m/z = 181).
It also breaks the symmetry of C-C stretching modes at 1567 and 1616 cm$^{-1}$.
The fragment (m/z = 181) has a dangling CH$_2$ group as a side group.
It is in the same plane as the aromatic ring and exhibits an aromatic C-H in-plane bending feature at 1300 cm$^{-1}$.
More band assignments of the fragment (m/z = 181) can be found in Sect. \ref{s-C2H4}.

  
\end{appendix}
\end{document}